\documentstyle[12pt]{article}
\title{ Logical Reversibility and Physical Reversibility \\
in Quantum Measurement
}

\author{Masahito Ueda\\
Department of Physical Electronics, Hiroshima University,\\
Higashi-Hiroshima 739, Japan}

\date{}

\begin{document}

\maketitle

\begin{abstract} 
A quantum measurement is {\it logically reversible} if the premeasurement
density operator of the measured system can be calculated from the
postmeasurement density operator and from the outcome of the measurement.
This paper analyzes why many quantum measurements are logically
irreversible, shows how to make them logically reversible, and discusses
reversing measurement that returns the postmeasurement state to the
premeasurement state by another measurement ({\it physical reversibility}).
Reversing measurement and unitarily reversible quantum operation are
compared from the viewpoint of error correction in quantum computation.
\end{abstract}

\section{Introduction}

In classical mechanics, one can, in principle, measure a quantity of an
object very precisely while keeping the backaction of the measurement as
small as one wishes. 
The situation drastically changes in quantum mechanics; the more precisely
one wants to measure a quantity of an object, the more strongly one has to
couple the measuring apparatus to the object with the greater backaction of
the measurement. 
In addition, the state reduction occurs in a way that depends on a specific
outcome of measurement which is unpredicable. 
A quantum measurement thus introduces an asymmetry in the direction of time. 
With respect to the past, it confirms the predicted probability distribution
of an observable by a number of measurements performed on an ensemble
representing the same quantum state. 
With respect to the future, it produces a new quantum state via state
reduction by a single measurement. 
Such disparate roles played  by a quantum measurement with respect to the
past and future are the origin of irreversibility in the {\it dynamical}
evolution of a quantum-mechanical system. 
I will nevertheless show that a broad class of quantum measurement can be
{\it logically reversible} in the sense that the premeasurement density
operator of the measured system can be calculated from the postmeasurement
density operator and from the outcome of measurement\,\cite{1}.

The concept of logical reversibility provides a criterion for deciding
whether or not the system's information is preserved in the course of
measurement. 
A quantum measurement is logically reversible if and only if all pieces of
information concerning the measured systems are preserved during the
measurement.
If some pieces of information are lost, the measurement is logically
irreversible. 

Recently, there has been considerable interest in restoring `lost' coherence
in a device that performs quantum computation because in a quantum computer
information is stored in a superposition of states and therefore decoherence
makes long computation impossible. 
One candidate to recover the `lost' coherence is unitarily reversible
quantum operation which restores the original state by a unitary
transformation, hence with unit probability\,\cite{2}. 
I will propose another possibility to resolve this problem, namely, by means
of a {\it reversing measurement} scheme\,\cite{3} and compare these two
methods. 

This paper is organized as follows. 
Section II analyzes why many familiar quantum measurements are logically
irreversible and examines conditions that are necessary for a measurement to
be logically reversible. 
Section III mathematically formulates logically reversible measurement and
describes some useful theorems. 
Section IV illustrates examples of logically reversible measurement. 
Section V discusses the necessary and sufficient condition for a measurement
process to be reversed by another measurement ({\it physical reversibility}). 
It will be shown, however, that such a {\it reversing measurement} scheme
cannot be used to measure the wave function of a single quantum system. 
Section VI compares a reversing measurement scheme with unitarily reversible
quantum operation from the viewpoint of error correction.

\section{Sources of Irreversibility in Quantum Measurement}

Let us analyze why many familiar quantum measurements are logically
irreversible and examine conditions that are necessary for a measurement to
be logically reversible.

Consider first a sharp position measurement of a massive particle, where
measurements are called sharp if they involve no measurement error, and
unsharp otherwise. 
Since the sharp position measurement belongs to Pauli's first-kind
measurement, the postmeasurement state $\psi(q,t^+)$ is uniquely specified
by the outcome $q_0$ of the measurement, i.e., $\psi(q,t^+)=\delta(q-q_0)$. 
This postmeasurement state, however, contains no information about the
premeasurement state except that the states before and after the measurement
have a nonzero overlap. 
An exceptional situation is the one in which the premeasurement state is
{\it a priori} known to be a position eigenstate; then the state is not
disturbed by the measurement, and therefore the outcome of the measurement
uniquely specifies both the postmeasurement state and the premeasurement state. 
Excluding such an exception, a sharp position measurement is logically
irreversible. 
In general, any sharp measurement is logically irreversible. Logically
reversible measurement should therefore be unsharp.

Unsharpness is necessary for a measurement process to be logically
reversible, but it is not sufficient. 
To understand this, let us consider photon counting. Photon counting is a
typical example of continuous measurement and has played a major role in the
theory of measurement. 
A unique feature of photon counting is that a photodetector does not
register all photons at a time, but rather one by one. 
In principle, an infinitely long observation time is necessary to uniquely
determine the number of photons enclosed in a cavity. 
Therefore, a photon counting process --- or, in general, any continuous
measurement process --- with a finite observation time can be viewed as
unsharp measurement. 
Photon counting consists of a time sequence of two fundamental processes,
namely, ``one-count'' and ``no-count'' processes\,\cite{4}. 
The one-count process means that one photon is detected during an
infinitesimal time, and the no-count process means that no photon is
detected during a time interval.

\section{Introduction}

In classical mechanics, one can, in principle, measure a quantity of an
object very precisely while keeping the backaction of the measurement as
small as one wishes. 
The situation drastically changes in quantum mechanics; the more precisely
one wants to measure a quantity of an object, the more strongly one has to
couple the measuring apparatus to the object with the greater backaction of
the measurement. 
In addition, the state reduction occurs in a way that depends on a specific
outcome of measurement which is unpredicable. 
A quantum measurement thus introduces an asymmetry in the direction of time. 
With respect to the past, it confirms the predicted probability distribution
of an observable by a number of measurements performed on an ensemble
representing the same quantum state. 
With respect to the future, it produces a new quantum state via state
reduction by a single measurement. 
Such disparate roles played  by a quantum measurement with respect to the
past and future are the origin of irreversibility in the {\it dynamical}
evolution of a quantum-mechanical system. 
I will nevertheless show that a broad class of quantum measurement can be
{\it logically reversible} in the sense that the premeasurement density
operator of the measured system can be calculated from the postmeasurement
density operator and from the outcome of measurement\,\cite{1}.

The concept of logical reversibility provides a criterion for deciding
whether or not the system's information is preserved in the course of
measurement. 
A quantum measurement is logically reversible if and only if all pieces of
information concerning the measured systems are preserved during the
measurement.
If some pieces of information are lost, the measurement is logically
irreversible. 

Recently, there has been considerable interest in restoring `lost' coherence
in a device that performs quantum computation because in a quantum computer
information is stored in a superposition of states and therefore decoherence
makes long computation impossible. 
One candidate to recover the `lost' coherence is unitarily reversible
quantum operation which restores the original state by a unitary
transformation, hence with unit probability\,\cite{2}. 
I will propose another possibility to resolve this problem, namely, by means
of a {\it reversing measurement} scheme\,\cite{3} and compare these two
methods. 

This paper is organized as follows. 
Section II analyzes why many familiar quantum measurements are logically
irreversible and examines conditions that are necessary for a measurement to
be logically reversible. 
Section III mathematically formulates logically reversible measurement and
describes some useful theorems. 
Section IV illustrates examples of logically reversible measurement. 
Section V discusses the necessary and sufficient condition for a measurement
process to be reversed by another measurement ({\it physical reversibility}). 
It will be shown, however, that such a {\it reversing measurement} scheme
cannot be used to measure the wave function of a single quantum system. 
Section VI compares a reversing measurement scheme with unitarily reversible
quantum operation from the viewpoint of error correction.

\section{Sources of Irreversibility in Quantum Measurement}

Let us analyze why many familiar quantum measurements are logically
irreversible and examine conditions that are necessary for a measurement to
be logically reversible.

Consider first a sharp position measurement of a massive particle, where
measurements are called sharp if they involve no measurement error, and
unsharp otherwise. 
Since the sharp position measurement belongs to Pauli's first-kind
measurement, the postmeasurement state $\psi(q,t^+)$ is uniquely specified
by the outcome $q_0$ of the measurement, i.e., $\psi(q,t^+)=\delta(q-q_0)$. 
This postmeasurement state, however, contains no information about the
premeasurement state except that the states before and after the measurement
have a nonzero overlap. 
An exceptional situation is the one in which the premeasurement state is
{\it a priori} known to be a position eigenstate; then the state is not
disturbed by the measurement, and therefore the outcome of the measurement
uniquely specifies both the postmeasurement state and the premeasurement state. 
Excluding such an exception, a sharp position measurement is logically
irreversible. 
In general, any sharp measurement is logically irreversible. Logically
reversible measurement should therefore be unsharp.

Unsharpness is necessary for a measurement process to be logically
reversible, but it is not sufficient. 
To understand this, let us consider photon counting. Photon counting is a
typical example of continuous measurement and has played a major role in the
theory of measurement. 
A unique feature of photon counting is that a photodetector does not
register all photons at a time, but rather one by one. 
In principle, an infinitely long observation time is necessary to uniquely
determine the number of photons enclosed in a cavity. 
Therefore, a photon counting process --- or, in general, any continuous
measurement process --- with a finite observation time can be viewed as
unsharp measurement. 
Photon counting consists of a time sequence of two fundamental processes,
namely, ``one-count'' and ``no-count'' processes\,\cite{4}. 
The one-count process means that one photon is detected during an
infinitesimal time, and the no-count process means that no photon is
detected during a time interval.

Let us first analyze the state evolution in the no-count process. The
density operator of the photon field immediately after the no-count process 
$\hat{\rho}(t+\tau)$ is related to that $\hat{\rho}(t)$ before it
by\,\cite{4}\,\cite{5} 
\begin{eqnarray}
\hat{\rho}(t+\tau) = \frac{\exp (-\frac{\lambda}{2} \hat{a}^{\dagger} 
\hat{a} \tau) \hat{\rho}(t) \exp (-\frac{\lambda}{2} \hat{a}^{\dagger}
\hat{a} \tau)}{{\rm Tr} [\hat{\rho}(t) \exp (-\lambda \hat{a}^{\dagger}
\hat{a} \tau)]} , \label{1}
\end{eqnarray}
where $\hat{a}^{\dagger}$ and $\hat{a}$ are the creation and annihilation
operators of the photon field, and $\lambda$ is a coupling constant between
the photon field and the photodetector. By inspection we find that
Eq.\,(\ref{1})  can be inverted, giving
\begin{eqnarray}
\hat{\rho}(t)=\frac{\exp (\frac{\lambda}{2} \hat{a}^{\dagger} \hat{a} \tau)
\hat{\rho} (t+\tau) \exp (\frac{\lambda}{2} \hat{a}^{\dagger} \hat{a}
\tau)}{{\rm Tr} [\hat{\rho}(t+\tau) \exp (\lambda \hat{a}^{\dagger} \hat{a}
\tau)]} . \label{2}
\end{eqnarray}
This formula gives the premeasurement density operator $\hat{\rho}(t)$ in
terms of the postmeasurement density operator $\hat{\rho}(t+\tau)$ and the
readout information that no photon has been detected during time $\tau$. The
no-count process is therefore logically reversible, though the state
evolution is nonunitary.

The density operator $\hat{\rho}(t^+)$ of the photon field immediately after
the one-count process is related to that $\hat{\rho}(t)$ before it
by\,\cite{4}\cite{5}
\begin{eqnarray}
\hat{\rho}(t^+)=\frac{\hat{a}\hat{\rho}(t)\hat{a}^{\dagger}}{{\rm
Tr}[\hat{\rho}(t)\hat{a}^{\dagger}\hat{a}]}. \label{3}
\end{eqnarray}
This formula cannot be inverted for $\hat{\rho}(t)$ because information
about the vacuum state vanishes upon operation of $\hat{a}$ (or
$\hat{a}^{\dagger}$) on $\hat{\rho}(t)$ from the left (or from the right). 
Physically, this means that the conventional photodetector does not respond
to vacuum field fluctuations and therefore $\hat{\rho}(t^+)$ does not
contain any information concerning the premeasurement vacuum state. 
The one-count process is therefore logically irreversible. 
From this example we find that sensitivity of the detector to the vacuum
state or, in general, sensitivity of the detector to all states in the
Hilbert space is essential for a measurement process to be logically reversible.

\section{ Mathematical Formulation of Logically Reversible Measurement}

The concept of logical reversibility in quantum measurement can be
formulated on a firm mathematical basis.

A theory of measurement should describe both the probability for each
outcome of measurement and the corresponding postmeasurement state.
For many cases of interest, these two roles are described by a family of
linear operators $\{ \hat{A}_{\nu} \}$\,\cite{Davies}. 
The probability for outcome $\nu$ is given by 
\begin{eqnarray}
{\rm Tr}[\hat{\rho}\hat{A}_{\nu}^{\dagger}\hat{A}_{\nu}], \label{new.1}
\end{eqnarray}
where $\hat{\rho}$ is the premeasurement density operator of the measured
system, and the postmeasurement state $\hat{\rho}_{\nu}'$ is given by
\begin{equation}
\hat{\rho}_{\nu} ' = \frac{\hat{A}_{\nu} \hat{\rho}
\hat{A}_{\nu}^{\dagger}}{ {\rm Tr} [\hat{\rho} \hat{A}_{\nu}^{\dagger}
\hat{A}_{\nu}] }. \label{4}
\end{equation}
Equation (\ref{4}) indicates that $\hat{A}_{\nu}$ plays a role of a
generalized projection operator.
The state change from the premeasurement density operator $\hat{\rho}$ to
the postmeasurement density operator $\hat{\rho}_{\nu}'$ may be viewed as a
mapping.
Let this mapping be denoted by $\Gamma_{\nu} :
\hat{\rho}_{\nu}'=\Gamma_{\nu}(\hat{\rho})$. In terms of $\Gamma_{\nu}$, the
concept of logical reversibility may be defined in the following way\,\cite{3}:
\begin{quote}
{\it Definition. If $\Gamma_{\nu}$ is a one-to-one mapping, a measurement
process described by an operator $\hat{A}_{\nu}$ is logically reversible. If
$\Gamma_{\nu}$ is a one-to-one mapping for every possible outcome $\nu$, the
entire measurement $\{ \hat{A}_{\nu} \}$ is logically reversible.}
\end{quote}
If $\Gamma_{\nu}$ is a one-to-one mapping, $\hat{A}_{\nu}$ has a left
inverse, {\it and vice versa}.
A measurement process described by an operator $\hat{A}_{\nu}$ is therefore
logically reversible if and only if $\hat{A}_{\nu}$ has a left inverse.
We can prove two useful theorems\,\cite{3}. The first one is\,\cite{3}
\begin{quote}
{\it Theorem 1. The necessary and sufficient condition for  a measurement
process with outcome $\nu$ to be logically reversible is that
$\hat{A}_{\nu}|\psi \rangle$ does not vanish for any nonzero vector $|\psi
\rangle$ in the Hilbert space.}
\end{quote}
This theorem implies that the outcome $\nu$ does not exclude the possibility
of any state for the premeasurement state. 
To put it another way, the detector responds to all states in the Hilbert space.
The second theorem is
\begin{quote}
{\it Theorem 2. For any sharp measurement $\{ \hat{A}_{\nu} \}$, there
exists a logically reversible measurement $\{ \hat{A}_{\nu}(\epsilon) \}$
that becomes arbitrarily close to $\{ \hat{A}_{\nu} \}$ as the measurement
error, which is characterized by $\epsilon$, approaches zero.}
\end{quote}
This theorem implies that a broad class of sharp (and hence logically
irreversible) measurements can be made logically reversible by making them
unsharp in such a manner that $\hat{A}_{\nu}(\epsilon)$ has a left inverse
and that a functional dependence of $\hat{A}_{\nu}(\epsilon)$ on $\epsilon$
is known. 
We will illustrate this theorem in the following section.

\section{ Examples of Logically Reversible Measurement}

Until now, four different  kinds of measurement have been shown to be
logically reversible. The first example is the so-called quantum
counter\,\cite{1}.
The quantum counter was proposed by Bloembergen as an infrared photon
detector\,\cite{6}. 
This device was later reconized by Mandel to be sensitive to vacuum field
fluctuations\,\cite{7}. 
The quantum counter is thus a vacuum-field-sensitive photon counter, and is
shown to perform a logically reversible measurement of photon number\,\cite{1}.

Subsequently, Imamo\=glu showed that a three-state $\Lambda$ atomic system,
which has been used to study electromagnetically induced transparency, can
be used for a logically reversible quantum nondemolition (QND) measurement
of photon number\,\cite{8}.

Royer then pointed out that an unsharp measurement of a spin-1/2 can be
reversed by another measurement with a nonzero probability of success\,\cite{9}.
This affords an example of {\it physical reversibility} in the sense that
the premeasurement state can be recovered from the postmeasurement state by
means of a physical process. 
We will return to this subject later.

Most recently, we have shown that an unsharp Kerr QND measurement of photon
number  is logically reversible\,\cite{3}.
The basic idea of this measurement is as follows.
Suppose that we measure the photon number of the signal light in a
nondestructive way.
For this purpose, we pass the signal light and the probe light through a
nonlinear medium called Kerr medium.
In this medium, information concerning the signal photon number is
transferred to the change in phase of the probe light.
Measuring this change by the homodyne detection, we can estimate the signal
photon number in a nondestructive way.

The state evolution in the Kerr medium is given by the unitary operator
$\hat{U}=\exp (i \kappa \hat{n}_{s} \hat{n}_{p})$, where $\hat{n}_{s}$ and
$\hat{n}_{p}$ are the photon number operator of the signal light and that of
the probe light, respectively, and $\kappa$ is an effective coupling
constant between the signal light and the probe light which is proportional
to the third nonlinear susceptibility and to the length of the medium.
When the probe light is initially in a coherent state with amplitude
$\alpha$, and when readout $\beta_{2}$ of the homodyne detection is
obtained, the generalized projection operator $\hat{A}$ which describes the
state change of the signal light according to Eq.\,(\ref{4}) is calculated
to be\,\cite{3}
\begin{equation}
\hat{A}(\nu,\epsilon)=\sum_{n=0}^{\infty} \sqrt{C_{n}(\nu, \epsilon)} |n
\rangle \langle n |, \label{5}
\end{equation}
where $|n \rangle$ is the  Fock state (i.e., the eigenstate of the photon
number), $\nu \equiv \beta_{2}/ |\alpha| \kappa$ gives an estimated photon
number and $\epsilon=1/(2\sqrt{ |\alpha| \kappa})$ gives the measurement error.
As long as the coupling constant $\kappa$ is finite, $\hat{A}$ is not a
sharp projection operator of the photon number but is distributed around an
estimated photon number $\nu$ according to the Gaussian distribution with
mean $\nu$ and width $\epsilon$ :
\begin{equation}
C_{n} (\nu, \epsilon)= \frac{1}{\sqrt{2 \pi \epsilon^{2}}} \exp \Big[
-\frac{(n- \nu)^2}{2 \epsilon^2} \Big] \label{6}
\end{equation}
As the intensity $| \alpha |^{2}$ of the probe light becomes infinite, the
measurement error $\epsilon$ becomes zero and therefore a measurement
process described by $\hat{A}$ approaches a sharp measurement of the signal
photon number: 
\begin{equation}
\lim_{\epsilon \to 0} C_n (\nu, \epsilon)=\delta(n-\nu).
\end{equation}
As long as the measurement error $\epsilon$ is nonzero, $C_{\mu}(\nu,
\epsilon)$ is always positive, so that $\hat{A} (\nu, \epsilon)$ has a left
inverse.
Therefore, an unsharp Kerr QND measurement is logically reversible.

\section{ Reversing Measurement and Impossibility of Measuring the Wave
Function of a Single Quantum System}

Logical reversibility implies an information preserving measurement.
It is not clear, however, if the premeasurement state can be recovered from
the postmeasurement state by another measurement which we will call {\it
reversing measurement}\,\cite{3}.
Let us discuss the necessary and sufficient condition for a measurement
process to have a reversing measurement and its implications.

Suppose that we perform a measurement $ \{ \hat{A}_{\nu} \} $ on a state
represented by density operator $\hat{\rho}$, and that an outcome $\nu$ is
obtained. 
The postmeasurement density operator $\hat{\rho}_{\nu}'$ is given by
Eq.\,(\ref{4}).
Let the  reversing measurement be $ \{ \hat{R}_{\mu}^{(\nu)} \} $, where
$\mu$ represents possible outcomes of this measurement, and let $\mu =0$ be
the ``successful outcome'', that is, if the outcome $\mu=0$ is obtained, the
postmeasurement state $\hat{\rho}_{\nu}''$ of this measurement is given by
the initial state $\hat{\rho}$.
The necessary and sufficient condition for such a successful reversal to
occur is that $\hat{R}_{\mu=0}^{(\nu)}$ satisfies
\begin{equation}
\hat{R}_{\mu=0}^{(\nu)} \hat{A}_{\nu} = c^{(\nu)} \hat{1} \label{7}
\end{equation}
where $c^{(\nu)}$ is a nonzero and finite c-number and $\hat{1}$ is the
identity operator. 
Because $\hat{R}_{\mu =0}^{(\nu)}$ should gives the probability of
successful reversal according to Eq.\,(\ref{new.1}), it has to be bounded. 
If Eq.\,(\ref{7}) holds, the postmeasurement density operator
$\hat{\rho}_{\nu}''$ corresponding to the outcome $\mu=0$ can easily be
shown to be equal to the initial state $\hat{\rho}$:
\begin{equation}
\hat{\rho}_{\nu}'' = \frac{\hat{R}_{\mu=0}^{(\nu)} \hat{\rho}_{\nu}'
\hat{R}_{\mu =0}^{(\nu)\dagger}}{{\rm Tr}[\hat{\rho}_{\nu}' \hat{R}_{\mu
=0}^{(\nu)\dagger} \hat{R}_{\mu =0}^{(\nu)} ] } = \hat{\rho}. \label{8}
\end{equation}
To put it another way, a specific outcome $\nu$ of the measurement $ \{
\hat{A}_{\nu} \} $ has a reversing measurement $ \{ \hat{R}_{\mu}^{(\nu)} \}
$ if and only if $\hat{A}_{\nu}$ has a bounded left inverse.

The existence of reversing measurement, {\it prima facie}, seems to imply
that the wave function of a single quantum system can be measured. The
arguement goes as follows. 
As a result of a sequence of two measurements we have two outcomes $\nu$ and
$\mu =0$, and the system has returned to its initial state [see
Eq.\,(\ref{8})]. 
By repeating the same sequence of measurements many times, one might think
that it is possible to measure the wave function of a single quantum system. 
If it were true, several consequences would result such as cloning of a
single quantum system and superluminal communication that should be impossible.

To resolve this apparent paradox, consider the joint probability that the
first measurement yields an outcome $\nu$ and that the successful reversal
$\mu =0$ occurs at the second measurement.
The joint probability of such successive measurements is given from
Eqs.\,(\ref{new.1}) and (\ref{7}) by 
\begin{equation}
{\rm Tr} [ \hat{\rho}(\hat{R}_{\mu =0}^{(\nu)} \hat{A}_{\nu})^{\dagger} (
\hat{R}_{\mu =0}^{(\nu)} \hat{A}_{\nu}) ] = | c^{(\nu)} |^2 . \label{9}
\end{equation}
The crucial observation is that the right-hand side is independent of the
state of the measured system $\hat{\rho}$.
This means that whenever a successful reversal occurs, we cannot obtain any
information about the initial state except for the information that the
initial state has a nonzero overlap with the state corresponding to the
outcome $\nu$. 
Because the existence of reversing measurement $ \{ \hat{R}_{\mu}^{(\nu)} \}
$ automatically guarantees that the first measurement $ \{ \hat{A}_{\nu} \}
$ is logically reversible, we should obtain {\it all} possible outcomes $
\nu_{1}, \nu_{2}, \nu_{3},\cdots $ by repeating the first measurement $ \{
\hat{A}_{\nu} \} $ followed by the reversing measurement $ \{
\hat{R}_{\mu}^{(\nu)} \} $ many times; but we cannot obtain the probability
distribution for these outcomes because, although the probability
distribution of each outcome $\nu$ or $\mu=0$ depends on the state of the
measured system, their joint probability distribution doesn't as seen from
Eq.\,(\ref{9}).

When the dimension of the underlying Hilbert space is finite, a linear
operator is always bounded.
Logical reversibility therefore guarantees the existence of reversing
measurement.
A simple example is an unsharp measurement of spin systems; the simplest
case of a spin-1/2 system was analyzed in Ref.\,\cite{9}.

When the dimension of the underlying Hilbert space is infinite, a logically
reversible measurement may not have a reversing measurement.
If so, we may introduce an approximate reversing measurement by truncating
the Hilbert space.
For example, consider an unsharp Kerr QND measurement of photon number. 
As we discussed in Sec.\,IV, as long as there is a nonzero measurement error
$\epsilon$, $C_{n}(\nu, \epsilon) > 0 $ for all outcomes $\nu$ and therefore
the generalized projection operator $\hat{A}(\nu, \epsilon)$ in
Eq.\,(\ref{5}) has a left inverse $\hat{B}(\nu, \epsilon)$:
\begin{equation}
\hat{B} (\nu, \epsilon) = \sum_{n=0}^{\infty } \frac{1}{\sqrt{C_{n} (\nu,
\epsilon)}} |n \rangle \langle n |.
\end{equation}
However, since $ 1/ \sqrt{C_{n} (\nu, \epsilon)} $ grows exponentially with
$n$ [see Eq.\,(\ref{6})], $\hat{B}(\nu, \epsilon)$ is not bounded.
In practice, however, we may truncate the Hilbert space at a finite value
$N$ of the photon number, and the operator thus defined
\begin{eqnarray}
\hat{B}'(\nu, \epsilon)= \sum_{n=0}^{N} \frac{1}{\sqrt{C_n (\nu, \epsilon)}}
|n \rangle \langle n|
\end{eqnarray}
is bounded, so that we may introduce an approximate reversing measurement $
\{ \gamma \hat{B}', \sqrt{\hat{1} - | \gamma |^{2} \hat{B}'{}^{\dagger}
\hat{B}'} \} $, where a nonzero c-number $\gamma$ is introduced to satisfy
the condition that the trace of $ | \gamma |^{2} \hat{B}'{}^{\dagger}
\hat{B}' $, which is the probability for successful reversal, is equal to or
less than unity.

\section{ Restoring Lost Coherence in Quantum Computation}

In a quantum computer, information is stored in a superposition of states. 
It is therefore crucial to protect the system against decoherence in order
to perform long computation.
Because the system is always subject to contact with its environment, the
system easily decoheres via, e.g., spontaneous emission.
Shor proposed to using redundant coding to overcome this problem\,\cite{Shor}.

A scheme of reversing measurement could provide another means to overcome
the problem of decoherence.
A key idea is that if a decoherence process may be regarded as logically
reversible measurement, we may construct (at least an approximate) reversing
measurement (see Sec.\,V) to restore the initial state.
A problem of this scheme is how to make the probability of successful
reversal close to unity.

If the system's initial states is known to lie within a special subspace of
the entire Hilbert space, we may restore the initial state by a unitary
transformation --- hence with unit probability\,\cite{2}.
Unitary transformation implies that there should be no information readout.
In fact, the subspace must be chosen such that within this subspace the
probability for each outcome is independent of the initial state and thereby
no information can be extracted about the initial state.
This scheme of restoring the initial state is called unitarily reversible
quantum operation\,\cite{Nielsen}.
The problem of this scheme is that it is not always easy to prepare an
initial state within such a restricted Hilbert space.

The two schemes --- reversing measurement and unitarily reversible quantum
operation --- have one common requirement.
That is, the decoherence process must occur in a logically reversible
manner; otherwise, some pieces of information about the initial state of the
system would be lost in that process and there would be no means to restore
the initial state.
If the initial state and the decohered state are known to be bounded as in
the case of unitarily reversible measurement, we can construct a number of
reversing measurements with varying probabilities of successful reversal.
The author conjectures that unitarily reversible quantum operation may be
viewed as a special case of reversing measurement at the limit of the
probability of successful reversal being equal to unity.

\begin{flushleft}
{\large {\bf Acknowledgments}}
\end{flushleft}

The author would like to thank N. Imoto and H. Nagaoka for fruitful
collaboration yielding Ref.\,\cite{3} on which parts of Sec.\,III-V are based.
This work was supported by the Core Research for Evolutional Science and
Technology (CREST) of the Japan Science and Technology Corporation (JST).

\end{document}